\documentclass[conference]{IEEEtran}
\IEEEoverridecommandlockouts
\usepackage{cite}
\usepackage{amsmath,amssymb,amsfonts}
\usepackage{algorithmic}
\usepackage{graphicx}
\usepackage{textcomp}
\usepackage[table,xcdraw]{xcolor}

\def\BibTeX{{\rm B\kern-.05em{\sc i\kern-.025em b}\kern-.08em
    T\kern-.1667em\lower.7ex\hbox{E}\kern-.125emX}}
\begin{document}

\title{Metamorphic Malware Evolution: The Potential and Peril of Large Language Models}

\author{\IEEEauthorblockN{Pooria Madani}
\IEEEauthorblockA{\textit{Business and Information Technology} \\
\textit{Ontario Tech University}\\
Oshawa, Canada \\
pooria.madani@ontariotechu.ca}

}

\maketitle

\begin{abstract}
Code metamorphism refers to a computer programming exercise wherein the program modifies its own code (partial or entire) consistently and automatically while retaining its core functionality. This technique is often used for online performance optimization and automated crash recovery in certain mission-critical applications. However, the technique has been misappropriated by malware creators to bypass signature-based detection measures instituted by anti-malware engines. However, current code mutation engines used by threat actors offer only a limited degree of mutation, which is frequently detectable via static code analysis. The advent of large language models (LLMs), such as ChatGPT 4.0 and Google Bard may lead to a significant evolution in this landscape. These models have demonstrated a level of algorithm comprehension and code synthesis capability that closely resembles human abilities. This advancement has sparked concerns among experts that such models could be exploited by threat actors to generate sophisticated metamorphic malware. This paper explores the potential of several prominent LLMs for software code mutation that may be used to reconstruct (with mutation) existing malware code bases or create new forms of embedded mutation engines for next-gen metamorphic malwares. In this work, we introduce a framework for creating self-testing program mutation engines based on LLM/Transformer-based models. The proposed framework serves as an essential tool in testing next-gen metamorphic malware detection engines.
\end{abstract}


\section{Introduction}
Code metamorphism refers to a programming methodology wherein the program self-modifies its code consistently and automatically, retaining its core functionality while changing its binary/syntactical representation. An intriguing application of code metamorphism lies in enhancing fault tolerance of mission-critical systems via code variant redundancy, allowing the system to switch to an automatically generated alternative version of the code when confronted with faults at runtime [1].

 The technique of code metamorphism has been adopted by creators of malicious software (i.e., malware) to circumvent detection by anti-malware applications [2]. Metamorphic malware can change its code (while preserving the functionality) at each iteration of replication, making signature-based detection, a gruelling feat. Often, this rewriting or reorganization of code can be carried out using rule-based code mutation engines through techniques, such as changing variable names, reordering instruction sequences, or using different programming constructs that achieve the same expected code functionality. These rule-based engines rely on a predefined array of templates to modify a given program code (or binary) into versions that are semantically equivalent, but syntactically distinct [2]. However, since the majority of contemporary code mutation engines lack the ability to interpret program code semantics, their internal rules permit only a relatively limited degree of mutation that is often easy to detect via static code analysis. This landscape is set to change significantly with the advent of large language models (LLMs) like ChatGPT 4.0 [3] and Google Bart [4] that are demonstrating human-level algorithm understandability and program code synthesis capabilities. Some researchers are concerned about threat actors potentially misusing emerging LLMs (capable of generating software code from English instructions) to develop and generate complex malwares.  

Large language models are types of artificial intelligence models that have been trained involving a colossal amount of web text data (e.g., news articles, blog posts, online source code repositories, etc.) and are capable of interpreting as well as generating human language contents (e.g., blog posts) in a way that is surprisingly coherent and contextually appropriate [5]. Additionally, it has been demonstrated that LLMs can assist with programming and coding tasks given that their training data includes substantial coding and technical discussions from various online sources [6]. However, it must be noted that LLMs do not truly understand code or programming principles and the quality of code they generate can be hit-or-miss, and they might produce code that is incorrect, inefficient, or insecure. 

In this paper, we present compelling arguments and substantiate them with experimental data where forthcoming LLMs hold immense potential as instrumental aids in constructing metamorphic computer programs. This powerful capability may very well be a double-edged sword. If misused, these models could serve as formidable tools in the creation of advanced metamorphic malware, posing a significant threat to the safety and stability of our increasingly interconnected world. Hence, our contribution is threefold:
\begin{enumerate} 
    \item Investigate the use of current LLMs models to regenerate software code while preserving existing semantics (i.e. demonstrating code mutation capabilities),
    \item proposed a framework to construct self-testing program mutation engines utilizing LLM/Transformer-based models,
    \item demonstrates the first proof-of-concept implementation of the proposed framework that utilizes few prominent open-source LLMs and OpenAI’s ChatGPT to achieve code metamorphism for any arbitrary Python codebase. 
\end{enumerate}

The remaining paper is organized as follows: Section 2 presents a review of currently established techniques in code metamorphism. In Section 3, we survey the existing literature on code synthesis using LLMs and evaluation metrics used in our study while outlining our LLM-based code mutation framework. Section 4 details the practical application of our framework, demonstrating its execution against some real-world Python programs. We conclude our discussion in Section 5, which addresses the limitations of our proposed approach and suggests potential areas for future research.

\section{EXISTING APPROACHES TO CODE METAMORPHISM}
Code metamorphism refers to a computer programming exercise wherein the program modifies its own code (partial or entire) consistently and automatically while retaining its core functionality [1]. This technique is often used for online performance optimization and automated crash recovery in certain mission-critical software. However, the technique has been misappropriated by malware creators to bypass detection measures instituted by anti-malware engines.

Signature-based detection anti-malware software operates by cross-examining potentially harmful files against a database of known malware signatures - unique code snippets that signify distinct malware [7]. The binary file under inspection is  designated malicious when it correlates with a malicious signature in the database. As such, code metamorphism emerges as a primary technique for malware creators to counteract signature-based anti-malware defences by continually altering the malware code (automatically and on the fly) to escape detection. Over time, malware authors have demonstrated significant innovation in devising new techniques for automatic code mutation (i.e., metamorphism), often staying one step ahead of existing defences. In this section, we are going to briefly discuss some of the most common code mutation techniques that are commonly used by malware authors.

\subsection{Instruction Substitution}
Instruction substitution is a technique in code metamorphism that constitutes certain instructions in a program’s code being replaced by another set of instructions while the overall functionality of the program remains intact. This technique assists in modifying the program code stored on disk without changing its overall behaviour, making it difficult to match the modified code against a database of known malicious code signatures. For instance, Figure 1 depicts three distinctly different Python statements possessing the same semantic – increasing the value of variable ‘a’ by 1. 

\begin{figure}[htbp]
\centerline{\includegraphics{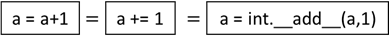}}
\caption{Three semantically identical Python expressions for increasing the value of variable ‘a’ by 1.}
\end{figure}

\subsection{Instruction Permutation}
Instruction permutation refers to the reordering of independent instructions in a program code while preserving its original functionality. In this context, independent instructions refer to any pair of instructions where an execution order does not affect the program’s overall functionality. Such reordering generates a different binary representation of the same program code, which in turn, makes it more difficult for signature-based detection systems to detect the presence of known malicious code sequences in the program. In addition, instruction permutation is effortlessly achieved in high-level programming languages, such as Python, as demonstrated in Figure 2.

\begin{figure}[htbp]
\centerline{\includegraphics{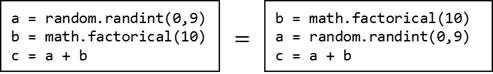}}
\caption{Two semantically identical Python code snippets demonstrating expression/instruction permutation.}
\end{figure}

\subsection{Variable/Register Substitution}
In the context of code metamorphism, variable/register substitution is the process of systematically replacing the variables/registers used in a program code with new ones. However, as many parts of a program code may be dependent on few variable/register identifiers, such changes must be made taking into consideration existing code dependencies on the modified variable/registers to ensure that the functionality of the code remains unchanged.

\begin{figure}[htbp]
\centerline{\includegraphics{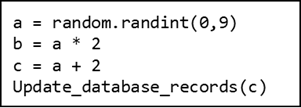}}
\caption{Python code snippet demonstrating dead code insertion – variable ‘b’ is never being used.}
\end{figure}

\subsection{Dead Code Insertion}
Dead code insertion involves the injection of non-functional (a code where its execution would not affect any important data or the original program’s functionality) code into a program’s source code. While not having any impact on the program output or its functionality, inserted dead codes can significantly change the program’s binary/source structure. For example, as depicted in Figure 3, the second statement does not influence the overall functionality of the program; it is still incorporated within the structure of the source code.

Identification of dead codes in a program source/binary (especially if dead codes contain a long sequence of function calls) is not a trivial task. It may require extensive static and dynamic code analysis that may be computationally prohibitive.

\subsection{Changing the Control Flow}
In the context of code metamorphism, threat actors can rearrange the sequence in which a program's individual statements or instructions are executed (by rearranging statements in a source code) while preserving its overall functionality and output. Undertaking such modifications necessitates a comprehensive understanding of the overarching program’s logic and proficiency in the programming language in which the program is written.

There are several control flow mutation techniques that are used to achieve code metamorphism:
\begin{itemize} 
    \item {\bf Unreachable code insertion:} code that cannot be reached during execution of the program, is inserted into the source code. This can confuse many static analysis tools.
    \item {\bf Control flow flattening:} the control flow of the program is modified to remove some of the hierarchies in the source code. 
    \item {\bf Jump instructions:} use of unnecessary jump instructions (supported by certain programming languages, such as Assembly, C, and Fortran) to create confusion about the control flows in the program source code.
\end{itemize}

Modifying the control flows used in a program would significantly change its source/binary structure. Determining equivalency between two versions of a program where the control flows are modified is an NP-Complete problem [8]. Therefore, control flow modification is considered a challenging yet highly effective technique for the creation of undetectable code metamorphism.

Undeniably, threat actors are not constrained by the techniques enumerated in this section. They are consistently innovating to craft increasingly sophisticated metamorphic malware. By combining the techniques discussed above, they can create malware variants that are extraordinarily difficult to detect, if not outright impossible, even by the most advanced anti-malware software of the day.

In the following sections, we will assess the efficacy of popular LLMs (capable of code synthesis) in facilitating code metamorphism using simple natural language instructions. We aim to study the potential of these models to serve as accessible tools to threat actors, enabling them to generate metamorphic malware without requiring extensive knowledge regarding the implementation of the discussed code mutation techniques in this section.

\section{CODE MUTATION USING LARGE LANGUAGE MODELS}
In machine learning, a generative model can learn the joint probability distribution of input features and labels (the training dataset), then uses this knowledge to create new data points that follow the same distribution as the training data [9]. This inherent capability enhances the utility of generative models in myriad application domains, such as speech synthesis, music generation, anomaly detection, code synthesis, and text generation/translation. 

Within the field of natural language processing (NLP), language models serve as a specific category of generative models that excel at learning underlying semantic and syntactic associations between words in a sequence of sentences. In turn, the trained language models are utilized to augment various NLP tasks, ranging from text synthesis and translation to sentiment classification. Over the last decade, different deep neural network architectures (e.g., recurrent neural networks) have demonstrated remarkable proficiency in language modelling and, thus, substantially outperformed classical methods, such as Hidden Markov Models (HMMs) [10] at the same tasks. In recent years, the term ``Large Language Model'' has emerged, alluding to the vast number of parameters trained within a deep neural network for language modelling purposes. This terminology underscores the pivotal role that neural networks play in the development and training of these expansive language models.  

In a seminal paper by Vaswani et al. [11] titled “Attention is All You Need,” a new deep neural network architecture called Transformers was introduced that revolutionized complex sequence modelling tasks specifically used in NLP, such as language modelling and text summarization. The proposed Transformer architecture relies on two novel mechanisms known as “Attention” and “Positional Encoding” to capture the importance and relationship of elements in sequential data in a parallel fashion. Leveraging their inherent capacity to process data in parallel, Transformers have drastically reduced the training time when compared to recurrent neural networks (RNNs) [12]. This advantage enables Transformer-based models to use exceptionally large datasets during training and achieve superior generalization capabilities when compared to their RNN-based counterparts.

Following the advent of Transformers, two models, namely Bidirectional Encoder Representations from Transformers (BERT) [13] and Generative Pretrained Transformers (GPT) [14] have emerged. These models have substantially enhanced the performance of various AI tasks and have thus, gained considerable recognition in the field. Developed by Google, BERT uses the Transformer’s Attention mechanism to analyze input sequence data in both directions (forward and backward in time) simultaneously, a method called bidirectional training. This enables BERT to understand the context of an element in a sequence based on all its surroundings (e.g., other words in a sentence) and learn a deeper understanding of the relationship between elements in a sequence regardless of where they appear. On the other hand, GPT developed by OpenAI, reads an input sequence from left to right and is trained to predict the next element in the sequence (i.e., auto-regression). 

Given the parallels between natural language documents and program source code, along with the demonstrated success of Transformers in creating original text content, it is intriguing to consider whether these same Transformer models could accomplish computer program synthesis, that is, the automated generation of program source code, provided they are trained on suitable data. In the remainder of this section, we survey some of the existing work in non-malicious computer program synthesis using Transformers and pre-trained large language models (LLMs) before presenting our framework for automatic code mutation using them.

\subsection{Computer Program Synthesis using Large Language Models (LLMs) }
Computer program synthesis, also known as automatic programming, is the act of creating programs that satisfy a certain set of specifications (e.g., natural language query or semantic similarity) with minimal human intervention and assistance. Program synthesis can assist with various tasks, such as automated code completion (when using a code editor software), generating bug fixes, optimizing existing code bases, code translation (between two programming languages), code clone detection, generating program code via a natural language description, or generating natural language comments for a program’s source code. Surprisingly, Transformer-based models are demonstrated to be effective with many of these tasks.

CodeBERT proposed by Feng et al. [15] is a bimodal BERT model, pre-trained for programming language and natural language tasks, such as natural language code search and code documentation generation. In the aforementioned work, authors demonstrated that BERT model can capture the semantic connections between natural language and programming language, and produce a general-purpose representation that can be useful for querying code while describing the program requirements using natural language. The authors demonstrated this capability over a wide range of programming languages, such as GO, Java, Python and Ruby, highlighting that learning program code semantics is not constrained to a particular programming language syntax. Furthermore, the demonstrated capacity of BERT models to understand the semantics of program codes holds promise for advancing other program synthesis tasks, such as automatic code mutation.

Codex created by Chen et al. [6] at OpenAI is another proprietary pre-trained (over their famous large language model) GPT model that can generate Python functions from natural language descriptions and docstrings with more than $70\%$ accuracy. For assessing the correctness of generated Python codes, authors resort to the use of pre-coded unit tests – an idea that has motivated the work conducted in this manuscript. Although their trained model is not open-source and not directly accessible to the public, users can interact with it through GitHub Codepilot and their famous ChatGPT prompt terminal.

CodeGen proposed by Nijkamp et al. [16] at Salesforce, strived to be an open-source version of Codex where natural language queries are turned into program code. Compared to Codex, CodeGen is generating far more inaccurate program codes. However, authors have demonstrated that by increasing the size of the language model parameters and incorporating more training data, their proposed generator can progressively create increasingly accurate outputs comparable to what is produced by Codex.  

Continual Pre-training on Sketches for Library-Oriented Code Generation (CERT) proposed by Zan et al. [17] is an open-source GPT-based automatic code generator developed at Microsoft. The authors articulated a two-stage generative system that can turn natural language instructions into Python codes while leveraging existing Python libraries, such as Pandas. For instance, when instructed to write a code to read a comma-separated value (CSV) file, CERT uses ``$pandas.read\_csv$'' function to accomplish the reading task instead of writing all the required code from scratch. They have trained their GPT model on 96GB high-quality Python codes extracted from GitHub that contain quality natural language comments while utilizing popular Python libraries, such as Pandas and NumPy in their bodies. 

AlphaCode proposed by Li et al. [18] at Google has tackled a much harder problem in program synthesizing. In contrast to the aforementioned works, which generally generate code snippets no larger than a function, AlphaCode can produce a complete program (potentially consisting of multiple functions), purely from a natural language description of the program's intended functionality. They begin by pretraining their Transformer-based language model with standard language model objectives (autoregression, i.e., given a sequence of $n$ tokens, generate the $n+1^{th}$ token). Then, they fine-tuned the model on Codecontest's competitive programming dataset in order to be able to answer programming competition-like queries and solve complex programming challenges. Authors have reported AlphaCode achieving on average, a ranking of top $54.3\%$ in competitions with more than 5,000 human participants. 

There are multiple other open-source projects that have trained LLMs for program code synthesis similar to the surveyed works above, summarized in Table 1.

\begin{table}[]
\caption{SUMMARY OF LARGE LANGUAGE MODEL CAPABLE OF CODE/PROGRAM SYNTHESIS}
\begin{center}
\begin{tabular}{|lrr|}
\hline
\multicolumn{1}{|l|}{\textbf{Model}}         & \multicolumn{1}{l|}{\textbf{Params}} & \multicolumn{1}{l|}{\textbf{\begin{tabular}[c]{@{}l@{}}HumanEval pass@1\end{tabular}}} \\ \hline
\multicolumn{3}{|c|}{\cellcolor[HTML]{9B9B9B}{\color[HTML]{FFFFFF} Closed Source}}                                                                                                \\ \hline
\multicolumn{1}{|l|}{GPT-3.5 {[}19{]}}       & \multicolumn{1}{r|}{175B}            & 48.10\%                                                                                     \\ \hline
\multicolumn{1}{|l|}{GPT-4 {[}3{]}}          & \multicolumn{1}{r|}{100T}            & 67.00\%                                                                                     \\ \hline
\multicolumn{1}{|l|}{Codex  {[}6{]}}         & \multicolumn{1}{r|}{12B}             & 28.80\%                                                                                     \\ \hline
\multicolumn{1}{|l|}{AlphaCode {[}18{]}}     & \multicolumn{1}{r|}{1.1B}            & 14.00\%                                                                                     \\ \hline
\multicolumn{1}{|l|}{Phi-1  {[}20{]}}        & \multicolumn{1}{r|}{1.3B}            & 50.6\%                                                                                      \\ \hline
\multicolumn{3}{|c|}{\cellcolor[HTML]{9B9B9B}{\color[HTML]{FFFFFF} Open Source}}                                                                                                  \\ \hline
\multicolumn{1}{|l|}{CodeGen-Multi {[}16{]}} & \multicolumn{1}{r|}{7B}              & \multicolumn{1}{c|}{-}                                                                      \\ \hline
\multicolumn{1}{|l|}{CodeGen-Mono {[}16{]}}  & \multicolumn{1}{r|}{16B}             & 29.3\%                                                                                      \\ \hline
\multicolumn{1}{|l|}{CodeGen2 {[}21{]}}      & \multicolumn{1}{r|}{16B}             & 19.1\%                                                                                      \\ \hline
\multicolumn{1}{|l|}{StarCoder {[}22{]}}     & \multicolumn{1}{r|}{15.5B}           & 33.6\%                                                                                      \\ \hline
\multicolumn{1}{|l|}{CodeT5+ {[}23{]}}       & \multicolumn{1}{r|}{16B}             & 30.90\%                                                                                     \\ \hline
\multicolumn{1}{|l|}{CodeParret {[}24{]}}    & \multicolumn{1}{r|}{1.5M}            & 3.58\%                                                                                      \\ \hline
\multicolumn{1}{|l|}{SantaCoder {[}25{]}}    & \multicolumn{1}{r|}{1.1B}            & 14.00\%                                                                                     \\ \hline
\multicolumn{1}{|l|}{WizardCoder {[}26{]}}    & \multicolumn{1}{r|}{16B}            & 57.3\%                                                                                     \\ \hline
\end{tabular}
\end{center}
\end{table}

\subsection{$variation@k$ and $pass@k$ Evaluation Metrics }
One predominant approach to assess the correctness/quality of a model-generated code is by matching it against a reference solution – a common practice adopted from the field of machine translation known as the Bilingual Evaluation Understudy (BLUE) score, introduced by Papineni et al. [27] at IBM Research. However, as explored by Ren et al. [28], unlike machine-generated natural language contents, some syntactically different codes can have identical semantics when executed. Consequently, the BLUE score needs to be heavily modified to not mistakenly penalize semantically correct generated mutated codes that look different syntactically (i.e., mutated). 

More recent works in program synthesis [29, 30] have turned to measuring the functional correctness of generated codes where a generated body of code is considered correct if it passes a set of unit tests. Kulal et al. [29] proposed to evaluate the capability of program synthesis of a model using the $pass@k$ metric where $k$ code samples are generated per each evaluation problem and a given problem is considered solved if any of the generated samples pass the associated unit tests. This metric reports the total fraction of problems solved. 

\begin{figure}[htbp]
\centerline{\includegraphics{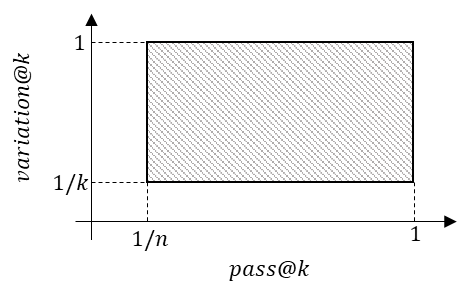}}
\caption{Region where LLM can be measured based on possible $pass@k$ and $variation@k$ values.}
\end{figure}

In this study, as opposed to the reviewed works in Section 3.A, the pass@k metric does not provide a dependable measure for the code mutation capability of LLMs. The issue we are addressing here involves not just the potential of a specific LLM to generate a syntactically accurate code snippet, but also its ability to produce multiple variations of codes for a given semantic. Hence, we introduce a new metric, $variation@k$ to measure the code mutation capability of a LLM-based model: for each problem $i$ in our evaluation dataset, $k$ code samples are generated and the average fraction of correct unique solutions (syntactically) per problem is calculated. Let $S=\{unique(s_1 )/k,…,unique(s_n )/k\}$ represent set of fractions of unique and correct solutions generated for the $n$ problems in the evaluation dataset using $k$ trials, then $variation@k$ is average of non-zero elements in $S$, defined as follows:

\begin{equation}
S' := \{ x|x\in S \& x >0\}
\label{eq:1}
\end{equation}

\begin{equation}
variation@k := \frac{1}{||S'||} \sum_{1}^{||S'||}s_i'
\label{eq:1}
\end{equation}

Figure 4 illustrates the defined regions and probable zones where a LLM-based model capability of code synthesis can be plotted based on its respective $pass@k$ and $variation@k$ metrics. It is important to highlight that not every area within the shaded region has an equal likelihood of occurrence. However, this graphical plot provides an effective foundation for comparing the performance of various LLMs in tasks related to code synthesis and code mutation.

\subsection{Proposed Framework for using LLMs in Code Mutation}
Often, a computer program comprises various subroutines or modules, which are executed in a systematic sequence as outlined by the software developer(s) within the program’s source code. Breaking down a program’s source code into smaller subroutines and modules helps with its readability, maintainability and testability. Each subroutine (e.g., function, procedure, class, etc.) can be tested in isolation (also known as unit testing) and its correctness can be verified independent of the remaining program. 

Even the slightest alteration to any of the program’s subroutines (in the source code) while maintaining the anticipated program semantics, will yield a new unique program source code. The modified source code can rightfully be deemed as a new variation of the original version. Moreover, as the alteration has occurred at the subroutine level, the correctness of modified subroutines can be verified through their associated unit tests. Therefore, in this work, we propose to use LLMs to modify subroutines in a program source code (e.g., malware source code) individually when creating new/different variations of the original source code.

\begin{figure}[htbp]
\centerline{\includegraphics{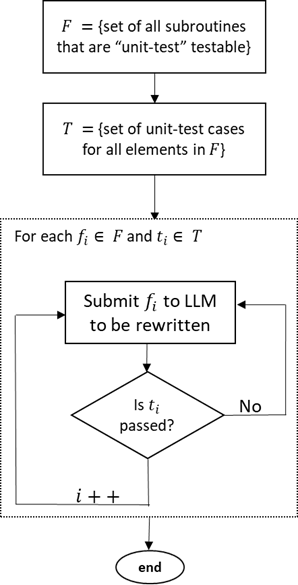}}
\caption{The proposed framework for using LLMs with ability of code/program synthesis to be used for source code mutation guided by unit test procedures.}
\end{figure}

We propose a framework specifically designed to assess the capabilities of Language Learning Models (LLMs) in the realm of code mutation. As depicted in Figure 6, our frameworks begins by enumerating all subroutines $f_i \in F$ in a source code that are associated with the unit-test $t_i$ (or unit test can be created for them in advance),  then iteratively submit each subroutine $f_i$ to the chosen LLM to be rewritten (i.e., mutated) and confirm the correctness of generated $f_i'$ against the unit-test $t_i$. 

In other words, instead of hoping for the entire code of a program to be rewritten by the chosen LLM correctly, the mutation process is broken down at the subroutine level and the correctness of modified subroutines are validated via associated unit tests.  Therefore, the ability of a LLM for program mutation boils down to its ability to generate multiple code versions of the constituent subroutines in a program source code. In the following section, we investigate the function-level (in Python programming language) mutation capability of several prominent LLMS to determine whether these models can effectively be utilized for code mutation following our proposed framework.

\section{Experimental Results}
The primary objective of this research is to scrutinize the code mutation prowess of some publicly accessible prominent LLMs that possess program synthesis capabilities. In order to validate our hypothesis and examine the effectiveness of our proposed framework, we have chosen 5 open-source LLMs along with OpenAI’s ChatGPT 3.5 as outlined in Table 2.

In our experiments, we have used OpenAI’s HumanEval dataset consisting of 164 hand-written programming problems along with several unit tests with “an average of 7.7 tests per problem” [6].  The programming tasks in this dataset are used to assess natural language comprehension, reasoning, algorithm developments and the simple mathematics of LLMs. What sets our experiments apart, a facet previously unexplored and unreported, is studying the capacity of the LLMs under scrutiny for the generation of  several distinct yet correct solutions to the given programming problems in this dataset.

It is important to recognize that the 6 models used in our study are not identical in their composition. These LLM models were initially trained for program synthesis tasks based on a series of NLP queries without specific fine-tuning for code mutation. The open-source models are all retrieved from Hugging Face (an online repository to access publicly shared LLMs) while ChatGPT 3.5 was queried through the public API provided by OpenAI.  

All LLMs used in our experiments are auto-regressive models, which means that they generate their output sequence one element at a time, where each new element (e.g., code symbols) is dependent on the previously generated elements. In other words, auto-regressive text/code (referred to as symbol and denoted as w) generation is based on the assumption that the probability distribution of a symbol sequence can be decomposed into the product of the conditional probability of the next symbol to appear as described in Equation 3.

 \begin{equation}
     P(w_{1:T}|W_0) := \prod_{t=1}^{T} P(w_t|w_{1:t} -1, w_0)
 \end{equation}

Thus, the probability of selecting a particular symbol (e.g., a keyword, a mathematical symbol, etc.) $w$ at time $t$ is proportional to the probability distribution  $w_t \sim P(w|w_{1:t-1})$. It is obvious that such text/code generation using sampling is not deterministic and each execution of this sampling process can potentially lead to generation of different outputs. In this work we have used a sampling scheme called Top-P proposed by Holtzman et al. [31] for which the smallest set of possible symbols (with k maximum number of symbols allowed to be in the set), the cumulative probability distribution exceeds the probability $p$ (considered at each time step). Following this, the probability mass is redistributed among this set of symbols before $w_t$ is drawn from the set. Of course, we have no control over the sampling process employed by OpenAI’s ChatGPT and rely on their black box model to generate the necessary outputs as we query their model remotely. 

We begin by querying each of the models under the study ten times to solve each of the programming tasks in the evaluation dataset. Then, each produced code snippet is parsed using Python’s AST library to remove human-readable comments and extra whitespace, and tested against provided unit test by HumanEval evaluation set to compute $pass@10$ score of each model. Furthermore, for each programming problem in the evaluation set, we have computed a SHA256 hash digest of each produced code snippet by any given model for identifying distinct synthesized codes and compiling variation@10 metrics. Table 2 summarizes the program synthesis performance and code mutation capabilities of the studied models using Top-P sampling (for the open-source models).  

As depicted in Figure 6 and Table II, in terms of code synthesis accuracy, ChatGPT is outperforming significantly all the other evaluated open-source models at $pass@10$. However, in terms of generating semantically similar code snippets that are syntactically different ($variation@10$), Salesforce’s open-source models are slightly lagging behind. It is important to note that none of the LLMs used in this study were explicitly trained to generate \emph{multiple unique} solutions to programming challenges. Nevertheless, they are exhibiting the ability to generate multiple variants of correct solutions for most of the presented programming challenges from HumanEval dataset. This observation can confirm our hypothesis that one can enhance the code synthesis mutability of these models significantly by specifically optimizing them to better cater to mutation-based objectives. 

\begin{table}[]
\caption{SUMMARY OF EVALUATION OF LLMS USED IN OUR CODE MUTATION STUDY. PASS@100 VALUES ARE SELF-REPORTED BY THE AUTHORS.}
\begin{center} 
\begin{tabular}{|l|r|r|r|}
\hline
\textbf{Models}            & \multicolumn{1}{l|}{\textbf{pass@10}} & \multicolumn{1}{l|}{\textbf{pass@100}} & \multicolumn{1}{l|}{\textbf{variation@10}} \\ \hline
CodeParrot {[}24{]}        & 8.03\%                                & 14.96\%                                & 5.32\%                                     \\ \hline
CodeGen2-Multi {[}21{]}    & 30.48\%                               & 50.80\%                                & 32.8\%                                     \\ \hline
CodeGen-Mono {[}16{]}      & 43.29\%                               & \textbf{75.00\%}                       & 38.45\%                                    \\ \hline
SantaCoder {[}25{]}        & 7.31\%                                & 49.00\%                                & 10.83\%                                    \\ \hline
StarCoderPlus {[}22{]}     & 10.36\%                               & \multicolumn{1}{c|}{-}                 & 11.76\%                                    \\ \hline
ChatGPT 3.5 Turbo {[}19{]} & \textbf{100.00\%}                     & \multicolumn{1}{c|}{-}                 & \textbf{51.32\%}                           \\ \hline
\end{tabular} 

\end{center}
\end{table}

The top three models demonstrated high complexity in generating different code snippets for some of the given programming challenges. For example, as depicted in Figure 7, CodeGen-Mono solved one of the programming challenges in two fundamentally different approaches: one using for-loop and the other using recursion. The demonstrated degree of variability between the two solutions is outside of the scope of techniques discussed in Section 2 and can hardly be replicated using rule-based mutation engines of the past.

\begin{figure}[htbp]
\centerline{\includegraphics{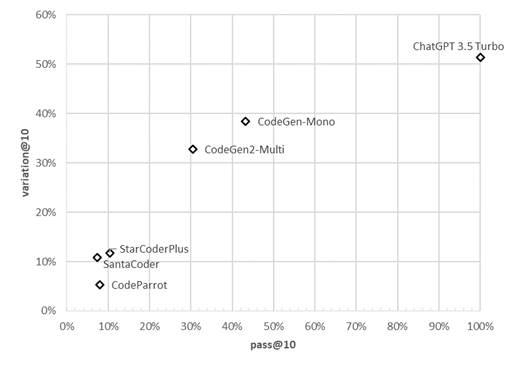}}
\caption{Region where LLM can be measured based on possible $pass@k$ and $variation@k$ values.}
\end{figure}

\begin{figure}[htbp]
\centerline{\includegraphics[width=\linewidth]{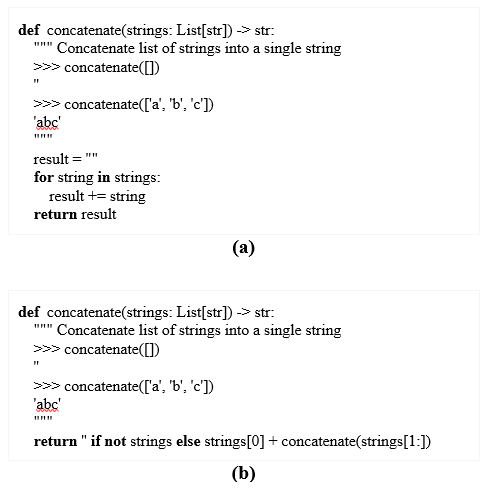}}
\caption{Two mutation examples generted by CodeGen-Mono-Python.}
\end{figure}

\section{Discussions and Future Work}
In this study, we have assessed the capabilities of several prominent large language models, both open and closed source in computer code synthesis and code mutation. Our findings unequivocally illustrate that current LLMs, even without explicit training for code mutation, possess the ability to generate diverse code snippets that are semantically identical. Consequently, the notion of explicitly training LLMs or constructing specialized Transformers for the task of code mutation is not an implausible concept. 

Undoubtedly, a pressing concern for existing large language models is based on their potential exploitation by threat actors to rewrite existing malware code bases and test the sensitivity of current anti-malware software to mutated malicious codes. Fortunately, the current large language models (LLMs) with code mutation capabilities, boasting billions of parameters, cannot be embedded in the core of metamorphic malware. The sheer size of these models renders them impractical for such usage (the top-performing models in our study have disk space requirements larger than 1 GB).  However, recent research [20] is pointing toward promising developments in the compression of large language models. These efforts aim to decrease the network size of LLMs while maintaining their synthesis capabilities, resulting in a reduction of disk space requirements. We hypothesize that near-future LLM/Transformer designs will enable the development of considerably smaller, yet highly capable code mutation engines. These advanced engines would possess the potential to be seamlessly integrated into malware binaries without significantly increasing the overall disk size of the malicious software. This hypothesis suggests a perilous future where more compact and potent AI-based code mutation engines can be employed by threat actors for the development of conspicuous malware. 

As an extension of this study, it would be crucial to conduct a similar investigation across various programming languages. The future research should aim to determine whether the grammatical structures of different programming languages impose constraints on code mutation and synthesis by LLMs and future code synthesis Transformers. Such a comprehensive study would enable the anti-malware development community to concentrate their efforts on a subset of malware codebases that utilize specific programming languages exhibiting high mutability potential through the utilization of forthcoming LLMs/Transformers.

Furthermore, we advocate for an in-depth study focused on training Transformer-based (other than LLMs) models for the purpose of code mutation. By training models explicitly for the code mutation task, we can streamline the evaluation loop of malware detection systems and effectively fortify our defences against the emergence of such cyber threats; it would equip us to address the advent of AI-based metamorphic malicious software proactively.

\end{document}